\documentclass{article}

\usepackage{PRIMEarxiv}

\usepackage[utf8]{inputenc} 
\usepackage[T1]{fontenc}    
\usepackage{hyperref}       
\usepackage{url}            
\usepackage{booktabs}       
\usepackage{amsfonts}       
\usepackage{nicefrac}       
\usepackage{microtype}      
\usepackage{lipsum}
\usepackage{fancyhdr}       
\usepackage{graphicx}       
\graphicspath{{media/}}     

\usepackage{cite}
\usepackage{amsmath,amssymb,amsfonts}
\usepackage{algorithmic}
\usepackage{graphicx}
\usepackage{algorithm,algorithmic}
\usepackage{hyperref}
\hypersetup{hidelinks=true}
\usepackage{textcomp}
\usepackage{algorithm}
\usepackage{array}
\usepackage[caption=false,font=normalsize,labelfont=sf,textfont=sf]{subfig}
\usepackage{stfloats}
\usepackage{booktabs}
\usepackage{url}
\usepackage{verbatim}
\usepackage{makecell}
\usepackage{color}
\usepackage{float}
\usepackage{multirow}
\usepackage{bm}
\usepackage{booktabs}

\title{An Adaptively Weighted Averaging Method for Regional Time Series Extraction of fMRI-based Brain Decoding
}

\author{
  Jianfei Zhu \\
  Faculty of Computing \\
  Harbin Institute of Technology \\
  Harbin, Heilongjiang, China\\
  \texttt{ } \\
   \And
  Baichun Wei \\
  School of Medicine and Health \\
  Harbin Institute of Technology \\
  Harbin, Heilongjiang, China\\
  \texttt{ } \\
   \And
  Jiaru Tian \\
  School of International Studies \\
  Harbin Institute of Technology \\
  Harbin, Heilongjiang, China\\
  \texttt{ } \\
     \And
  FengJiang \\
  School of Medicine and Health \\
  Harbin Institute of Technology \\
  Harbin, Heilongjiang, China\\
  \texttt{fjiang@hit.edu.cn} \\
     \And
  Chunzhi Yi \\
  School of Medicine and Health \\
  Harbin Institute of Technology \\
  Harbin, Heilongjiang, China\\
  \texttt{chunzhiyi@hit.edu.cn} \\
}
\begin{document}
\maketitle

\begin{abstract}
Brain decoding that classifies cognitive states using the functional fluctuations of the brain can provide insightful information for understanding the brain mechanisms of cognitive functions. Among the common procedures of decoding the brain cognitive states with functional magnetic resonance imaging (fMRI), extracting the time series of each brain region after brain parcellation traditionally averages across the voxels within a brain region. This neglects the spatial information among the voxels and the requirement of extracting information for the downstream tasks. In this study, we propose to use a fully connected neural network that is jointly trained with the brain decoder to perform an adaptively weighted average across the voxels within each brain region. We perform extensive evaluations by cognitive state decoding, manifold learning, and interpretability analysis on the Human Connectome Project (HCP) dataset. The performance comparison of the cognitive state decoding presents an accuracy increase of up to 5\% and stable accuracy improvement under different time window sizes, resampling sizes, and training data sizes. The results of manifold learning show that our method presents a considerable separability among cognitive states and basically excludes subject-specific information. The interpretability analysis shows that our method can identify reasonable brain regions corresponding to each cognitive state. Our study would aid the improvement of the basic pipeline of fMRI processing.
\end{abstract}

\keywords{Brain decoding \and deep learning \and regional time series\and task fMRI}

\section{Introduction}
For decades, how the brain processes information through segregation and integration is one of the key questions in neuroscience\cite{shine2019neuromodulatory,zuberer2021integration}, relating to identifying the brain regions that present specific activation or interaction patterns during specific cognitive tasks\cite{holland2001normal,tong2012decoding,zarahn2007age}. The advancement of modern neuroimaging techniques, such as fMRI, enables probing the brain activation corresponding to cognitive tasks in vivo. Traditionally, researchers utilized the general linear model as a tool to analyze how certain brain regions were activated when the subjects were performing some cognitive tasks, e.g., working memory (WM)\cite{dores2017study,narayanan2005role}, decision making\cite{esposito2009combined,rebola2012functional}, and motor learning\cite{macintosh2007brain,niu2021modeling} etc. Such analyses provided an inference of mapping the cognitive states of the brain with brain activation patterns. The inverse problem that classifies cognitive states by the brain activation patterns is called the “brain decoding” problem\cite{haynes2006decoding}. By decoding the cognitive states, some algorithms provided valuable information on how the function and computation of the brain work during cognition\cite{ito2020discovering,mensch2017learning,varoquaux2018atlases} and some cognitive impairments\cite{kapogiannis2011disrupted}. The feasibility of brain decoding is well demonstrated in the literature\cite{li2021braingnn,ye2023explainable,zhang2021functional}, and increasingly provides a useful tool for understanding the organization of the brain during cognition.

Among most of  the brain decoding studies, there are some common procedures, including data preprocessing, brain parcellation, regional time series extraction, and decoding algorithm development. Typically, node-based methods extract regional time series by averaging the voxel-wise signals within each region of interest (ROI)\cite{li2021braingnn,saeidi2022decoding,ye2023explainable,zhang2021functional}, after parcellating the 4-dimensional (4D) volumes of fMRI into brain regions (i.e. ROIs). Such a spatial average results in a 1D regional time series denoting the brain activity of the region, decreasing the data dimension to save computational requirements, has been thought of as following the functional segregation principle of brain organization\cite{friston2011functional}. Compared to voxel-wise approaches\cite{cheung2023decoding,kriegeskorte2006information,koelsch2021neocortical}, averaging voxels within each ROI can reduce computational requirements, enabling more intricate modeling of fMRI data. However, the spatial average treats every voxel within an ROI equally, which may neglect the spatial information that reflects voxel-wise signal distribution within an ROI. For example, Vu et al.\cite{vu2020fmri} conducted a study in which they applied a 3D convolutional neural network to classify raw fMRI volumes into different experimental conditions. The feature maps capture information at the voxel level for recognizing experimental conditions. Zhao et al.\cite{zhao2019four} also presented the varied voxel-wise activation within brain regions. These findings suggest that voxel-wise signal distribution within an ROI carries valuable information. Moreover, from the perspective of compressed sensing, image or signal compression should consider the downstream tasks, in order to retain the task-specific information and benefit the downstream tasks\cite{rani2018systematic}. The spatial average within ROIs extracts the 1D regional time series in a task-independent manner, without considering the downstream tasks of cognitive state decoding. Thus, adaptively weighted extraction of spatial information within each ROI, aligning with the requirements of downstream tasks, would be beneficial.

Regarding the downstream tasks, current solutions to the brain decoding problem generally involve deep learning-based methods. Such methods identify the cognitive states of the brain using the windowed regional time series of all the ROIs. In light of feature selection or explainable decoding algorithms, we can analyze the specific brain activation patterns corresponding to each cognitive state directly and avoid using the statistical analysis-based methods that identify the brain patterns by significance\cite{ye2023explainable}. In this way, some subtle activation patterns across the brain can be found\cite{woo2017building}. The deep learning-based methods utilize task fMRI scans that involve subjects experiencing various cognitive task conditions as input. Various models such as Graph Neural Networks (GNN)\cite{li2021braingnn,ye2023explainable,zhang2021functional,li2023multi}, Convolutional Neural Networks (CNN)\cite{ngo2022predicting,vu2020fmri,zhang2023unsupervised} and attention-based neural networks\cite{jiang2022attention} have been applied. These models have demonstrated satisfactory performance in cognitive state decoding. Notably, the performance of cognitive state decoding is influenced not only by the model design but also by the input signal. Factors such as the extent of image preprocessing\cite{vu2020fmri} and the granularity of atlases in brain region parcellation\cite{ye2023explainable} have been found to impact cognitive state decoding performance. Therefore, to validate the effectiveness of adaptively preserving spatial information, we can examine its impact on cognitive state decoding performance. In this study, we design a cognitive state decoding algorithm that learns the spatiotemporal information of the brain dynamics on two atlases with different granularity and use the decoder as the test bed to evaluate the effectiveness of adaptively weighting the voxels within each ROI. 

\begin{figure*}[ht]
\centering
\includegraphics[scale=.6]{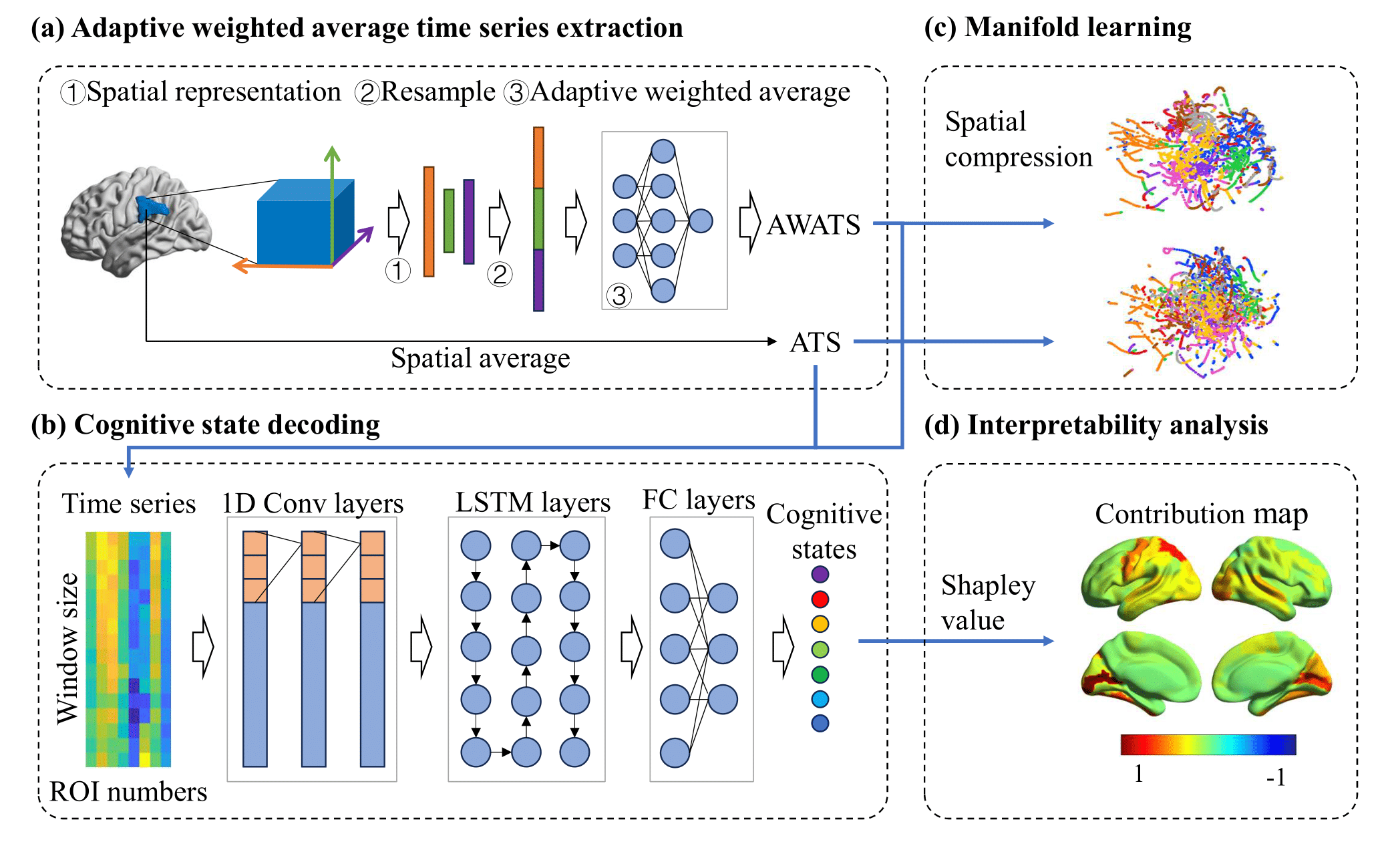}
\caption{Overview of the workflow of this paper. (a) The AWATS extraction method measures the data of a single ROI and of a TR from x, y and z axes, generating three spatial representation vectors. Then, the representation vectors are resampled to the same length and adaptively averaged by a fully connected network that is jointly trained with the downstream tasks (i.e. cognitive state decoding). (b) The structure of the STDfMRI model is designed for comparing AWATS and ATS in cognitive state decoding. (c) AWATS and ATS are spatially embedded into a two-dimensional space for temporal trajectory visualization. (d) Interpretability analysis is performed for the STDfMRI model using the Shapley value method and the ROI contributions are presented as brain maps.}
\label{fig:workflow}
\end{figure*}

In this study, we propose a novel Adaptively Weighted Average Time Series (AWATS) extraction method that aims to preserve the spatial information of voxel-wise signal within each ROI and extracting the information for the downstream tasks when extracting regional time series from the fMRI data. As shown in Figure \ref{fig:workflow}, three spatial representation vectors for all the voxels of an ROI at each TR (repetition time) are first generated by computing the means along the three axes (i.e. x, y, and z). Subsequently, these three spatial representation vectors are adaptively averaged using a fully connected neural network, which is jointly trained with the downstream task, to obtain the AWATS. Then, we validate the advantage of AWATS compared to the traditional Average Time Series (ATS) from two perspectives including cognitive state decoding and manifold learning. For the manifold learning, we perform a visualization analysis that embeds AWATS and ATS obtained from the task fMRI data into a two-dimensional space. For the cognitive state decoding, we design the Spatial-temporal Deep fMRI (STDfMRI) model for cognitive state decoding. This model takes AWATS or ATS as the input to predict the cognitive state, and the performances achieved using these two inputs are compared. In this way, we compare the distribution and the separability of the two types of regional time series data. Then, we perform an interpretability analysis for the cognitive state decoding. We utilize the Shapley value method to investigate which ROIs play an important role in identifying the cognitive state. Our study makes the following key contributions:

\begin{enumerate}
\item To the best of our knowledge, this study is the first work that proposes to replace the traditional averaging manner by adaptively weighting the voxels within each ROI in regional time series extraction, which would be valuable for various downstream tasks in fMRI computational analysis;
\item We evaluate the benefits of our method AWATS over ATS by performing extensive comparisons on the results of cognitive state decoding and manifold learning;
\item We perform Shapley value analysis to demonstrate the explainability of AWATS and provide some physiological insights into cognitive states decoding.
\end{enumerate}

\section{Materials and methods}
\subsection{fMRI data acquisition and preprocessing}
In this study, we analyze task-based fMRI data obtained from the HCP S1200 release dataset\cite{barch2013function}. The HCP S1200 release dataset comprises a total of about 1200 subjects, each with both resting-state and task fMRI measurements. HCP provides ethical approval with the dataset and no additional institutional review board (IRB) approval is required. The task fMRI includes seven tasks: emotion, gambling, language, motor, relational, social, and working memory (WM). Each task involves participants experiencing different task conditions, resulting in a total of 23 task conditions across the seven tasks. We use the 23 task conditions as the cognitive states and labels. The HCP dataset provides minimally preprocessed fMRI data\cite{glasser2013minimal}, which has already undergone motion correction and spatial normalization to the Montreal Neurological Institute's (MNI) standard space.

In this paper, we adopt the Schaefer atlases, which consider functional relevance when parcellating brain regions\cite{schaefer2018local}. In order to validate our method across atlases with different resolutions, we use Schaefer atlases at various scales (100, 200, 300, and 400 ROIs). We compute the average signals across voxels within each ROI, and denote the averaged regional time series as ATS.

\subsection{AWATS extraction}
\begin{figure}[ht]
\centering
\includegraphics[scale=.6]{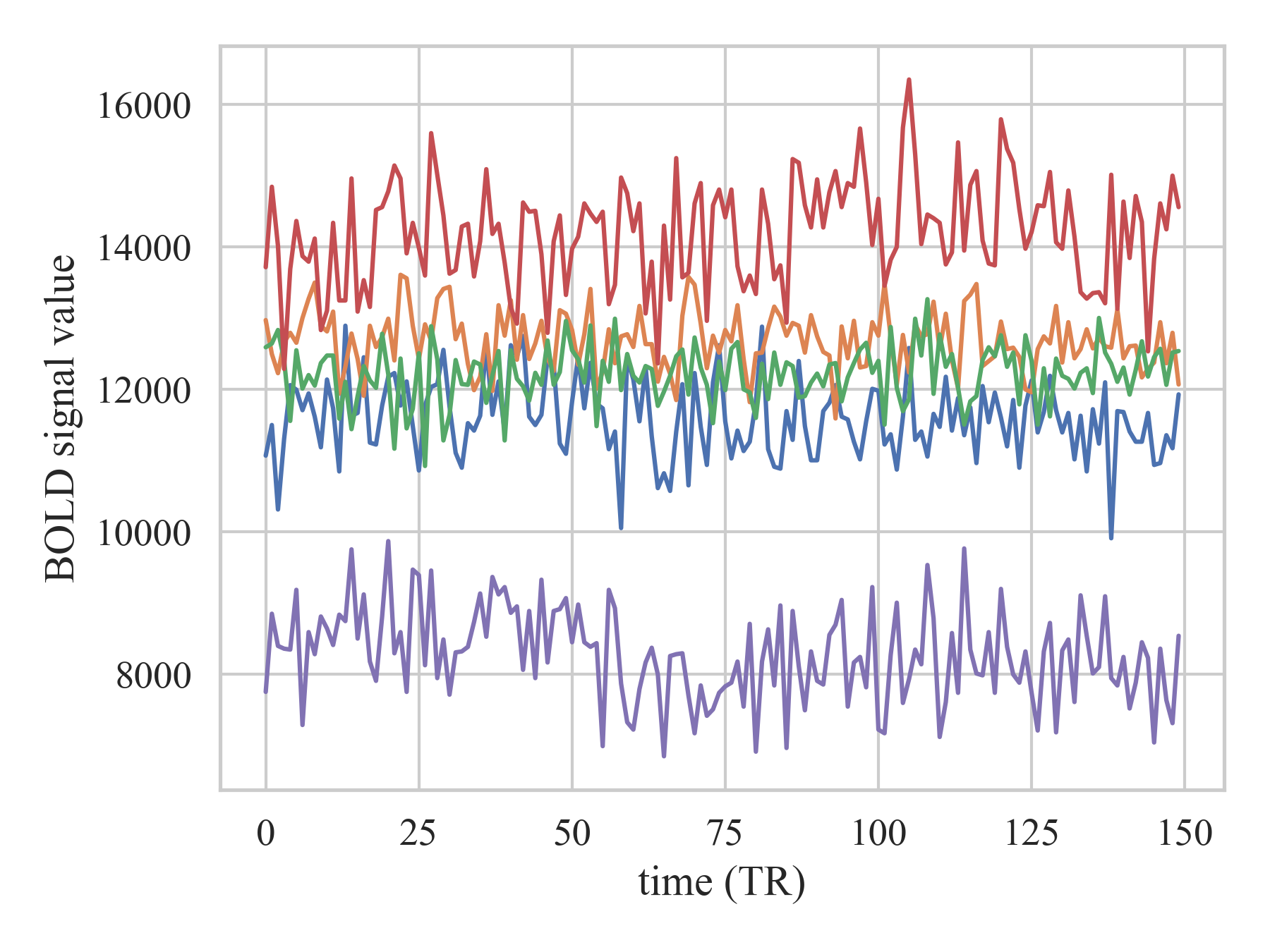}
\caption{BOLD signals of 5 voxels from the fist ROI of Schaefer 100.}
\label{fig:voxel_signals}
\end{figure}

The traditionally used ATS calculates the average value across the voxels of each ROI as its neural activities to extract regional time series from 4D fMRI data. This averaging method may neglect the spatial information with specific brain areas. To illustrate the potential spatial information within ROIs, we select the fMRI data of the first ROI of the Schaefer 100 atlas during the WM task as an example. Then, we present how the blood-oxygen-level-dependent (BOLD) signals of 5 voxels sampled from the first ROI varied across various TRs. Figure \ref{fig:voxel_signals} demonstrates that even within a single ROI, there exists variability in BOLD signal values among different voxels, and these values also change over time. This indicates that considering the spatial and temporal dynamics within each brain region may provide valuable insights beyond the traditional averaging approach. We introduce the AWATS extraction method, composed of three sequential steps: spatial representation, representation vector resampling, and adaptive weighted averaging.

\subsubsection{Spatial representation for each ROI}
In general, a single ROI typically comprises a substantial number of voxels and exhibits an irregular shape. We aim to use a neural network (NN) to learn to extract the AWATS. If we put all the voxels within each ROI into a NN, the memory and computational costs would be high and the different numbers of the voxels within ROIs would complicate the implementation of extracting the AWATS from each ROI. Therefore, as the first step, we generate three representation vectors by calculating the mean of the ROI from three different directions in each TR. In this way, the data of all voxels within the ROI and of each TR can be represented in a lower dimension while preserving spatial information.

Specifically, let $\mathbf{M}_i \in \mathbb{R}^{X \times Y \times Z}$ (X, Y, and Z denote the size of a single fMRI volume in three dimensions) represent the mask for the i-th ROI:

\begin{equation}
\mathbf{M}_i(x, y, z) =
\begin{cases}
  1, & \text{the voxel }(x, y, z) \text{ in the } i\text{-th ROI} \\
  0, & \text{else}
\end{cases}
,
\end{equation}

\noindent
then the volume of the i-th ROI can then be represented as:

\begin{equation}
\mathbf{F}_i = \mathbf{F} \odot \mathbf{M}_i,
\end{equation}

\noindent
where $\odot$ represents the element-wise multiplication of matrices, $F\in \mathbb{R}^{X \times Y \times Z}$ represents the fMRI data of a single TR, and $\mathbf{F}_i$ represents the fMRI data of a single TR specific to the i-th ROI. Indeed, the data from a single ROI may not cover the range of the whole brain. To focus on the relevant data, we subsequently clip both $\mathbf{F}_i$ and $\mathbf{M}_i$. This involves retaining solely the data range deemed relevant by identifying the corresponding indices of the useful data:

\begin{equation}
\mathbf{I}_{i,X} = \text{find}\left(\sum_{y, z} \left[\mathbf{M}_i(x, y, z) > 0\right]\right),
\end{equation}

\noindent
where $\text{find}(\cdot)$ returns indices that satisfy the given condition, and $\mathbf{I}_{i,X}$ represents the indices for the X-axis where data from the i-th ROI exists. Similar calculations are performed for $\mathbf{I}_{i,Y}$ and $\mathbf{I}_{i,Z}$. We then apply this clipping to both $\mathbf{F}_i$ and $\mathbf{M}_i$ to obtain:

\begin{equation}
\mathbf{F}_i' = \mathbf{F}_i \left[\mathbf{I}_{i,X}, \mathbf{I}_{i,Y}, \mathbf{I}_{i,Z}\right];
\end{equation}

\begin{equation}
\mathbf{M}_i' = \mathbf{M}_i \left[\mathbf{I}_{i,X}, \mathbf{I}_{i,Y}, \mathbf{I}_{i,Z}\right].
\end{equation}

\noindent
here, we use $[\cdot, \cdot, \cdot]$ to represent indexing the matrix of specific indices, and $\mathbf{F}_i' \in \mathbb{R}^{X' \times Y' \times Z'}$ and $\mathbf{M}_i' \in \mathbb{R}^{X' \times Y' \times Z'}$ represents the fMRI data and the mask of the $i$-th ROI after clipping, where $X'$, $Y'$, and $Z'$ represent the size of the clipped matrix along the three axes corresponding to the length of $\mathbf{I}_{i,X}$, $\mathbf{I}_{i,Y}$ and $\mathbf{I}_{i,Z}$ respectively. Our goal is to create a representation that describes the spatial distribution of ROI data. To achieve this, we calculate mean values along the $X$, $Y$, and $Z$ axes, resulting in three representation vectors:

\begin{equation}
\mathbf{v}_x(x) = \frac{\sum_{y,z} \mathbf{F}_i'(x, y, z)}{\sum_{y,z} \mathbf{M}_i'(x, y, z)};
\end{equation}

\begin{equation}
\mathbf{v}_y(y) = \frac{\sum_{x,z} \mathbf{F}_i'(x, y, z)}{\sum_{x,z} \mathbf{M}_i'(x, y, z)};
\end{equation}

\begin{equation}
\mathbf{v}_z(z) = \frac{\sum_{x,y} \mathbf{F}_i'(x, y, z)}{\sum_{x,y} \mathbf{M}_i'(x, y, z)}.
\end{equation}

\subsubsection{Resampling the representation vectors}
Since the ROIs typically exhibit irregular shapes and different ROIs exhibit variations in data size, leading to varying lengths of the representation vectors $\mathbf{v}_x$, $\mathbf{v}_y$, and $\mathbf{v}_z$, we employ a linear interpolation algorithm to resample the representation vectors.
For the representation vector $\mathbf{v}$ with $p$ elements:

\begin{equation}
\mathbf{v} = [v_1, v_2, \ldots, v_p],
\end{equation}

\noindent
we aim to generate a new representation vector $\mathbf{v}'$ with $q < p$ elements. We denote $v_i'$ as the $i$-th element in $\mathbf{v}'$. Its corresponding location in $\mathbf{v}$ will be $\frac{i \cdot p}{q}$. We then identify neighbors for $v_i'$ in $\mathbf{v}$, denoting them as $v_j$ and $v_k$, satisfying $j < \frac{i \cdot p}{q} < k$. Subsequently, $v_i'$ can be represented as:

\begin{equation}
v_i' = v_j + \frac{v_k - v_j}{(k - j)}\left(i \cdot \frac{p}{q} - j\right),
\end{equation}

\noindent
where $j$ and $k$ are integers that are closest to $\frac{i \cdot p}{q}$. This algorithm is utilized to resample each representation vector into a standardized dimensionality, ensuring uniform representation dimensions across all ROIs. In this study, the resampling size $q$ is set to 10. In the following comparisons, we will conduct experiments to investigate the impact of a smaller resampling size on cognitive state decoding.

\subsubsection{Adaptively weighted average using a fully connected NN}
Following the above steps, we obtain three resampled representation vectors for each ROI and for each TR, characterizing the spatial distribution of the fMRI data within the ROI. We herein employ a fully connected NN as a feature extractor to extract the Adaptive Weighted Average Time Series (AWATS) from the resampled representation vectors. In this way, we expect to adaptively preserve the spatial information within each ROI.

Each feature extractor comprises a single fully connected hidden layer with 128 neurons followed by a ReLU activation function and batch normalization. The output layer of the feature extractor contains a single neuron. To create a unified input vector, we concatenate the three representation vectors corresponding to each ROI. So, the input layer of the module contains $3q$ neurons. When the feature extractor takes the three representation vectors of all TRs as the input, the input has a dimensionality of $3q \times T$, where $T$ corresponds to the number of the time points, while the output has a dimensionality of $1 \times T$. The feature extractor is jointly trained with downstream tasks (i.e. decoding the cognitive states). When training the model, the feature extractors learn to automatically extract the regional time series from the spatial representation vectors of all the TRs.

\subsubsection{Cognitive state decoding using STDfMRI}
To compare the differences in dynamic information between AWATS and ATS, we develop a deep neural network model called Spatial-temporal Deep fMRI (STDfMRI) for decoding the cognitive states. In the following, we adopt the task fMRI data from the HCP dataset and use the time window method to segment the time series into blocks as the task conditions. STDfMRI takes the  AWATS and ATS calculated by the windowed  time series data of each
voxel as the input, respectively, and maps them to the brain cognitive states.

\paragraph{Time windowing}

In each run of the task fMRI data acquisition from the HCP dataset, participants engage in a specific task and undergo different task conditions, leading to 23 cognitive states. To realize cognitive state decoding, the fMRI data from each run is segmented into several data blocks, with each block corresponding to a specific task condition. Task conditions are recorded as events, which contain the onset and duration of the respective task condition. We position the middle of the task condition and utilize a time window with the same midpoint to segment a data block corresponding to the task condition.

Given the duration of the task conditions and the intervals between adjacent conditions, we determine that the length of the time window is set to 15. In the following comparisons, we will conduct experiments to investigate the impact of shortening the time window size in cognitive state decoding.

\paragraph{The STDfMRI model architecture}
The STDfMRI model is illustrated in Figure \ref{fig:workflow}(c). The model is designed to map the extracted regional time series (AWATS or ATS) with the cognitive states, i.e. the task conditions. STDfMRI comprises a convolutional block, an LSTM block, and a fully connected block. The convolutional block is to extract features from the regional time series data in both the spatial and the temporal domains, while the LSTM block is designed to capture the temporal relationships in the regional time series data. The convolutional block comprises three 1D convolutional layers, each with 128 filters, and a kernel size of 3 with the padding set to 1 to maintain the input sequence length. The LSTM block consists of three layers with a hidden size of 128. Following the LSTM block, the fully connected block has a hidden layer with 64 neurons and an output layer with 23 neurons, corresponding to the number of HCP task conditions.

\paragraph{Experiment setting}
When using AWATS as the input of the STDfMRI model, the feature extractor in the AWATS extraction method is jointly trained with the STDfMRI model. In this way, AWATS can be automatically extracted from the sampled representation vectors.
The model is trained on an NVIDIA GeForce RTX 3070 GPU. During the training of the STDfMRI model, hyperparameters are adjusted to achieve optimal performance. Cross entropy is employed as the loss function, and the model is optimized using the Adam optimizer. The training is executed for 100 epochs, with a batch size of 128 and a learning rate of 0.0001. Our model undergoes a rigorous evaluation process involving training, validation, and testing to avoid a potential risk of leakage. Specifically, 80\% of the data is allocated for training, while 10\% each is allocated for validation and testing.  This procedure is iterated ten times, with the dataset randomly partitioned each time. This ensures a thorough assessment across diverse dataset divisions, contributing to the robustness of our findings. 

\subsection{Manifold learning for fMRI regional time series data}
The human brain is composed of a vast number of neurons, resulting in highly intricate brain dynamics. Brain activity measurements are typically represented as high-dimensional data. However, many studies have indicated that the complex dynamic changes and interactions in brain activity can be abstracted as brain states in low-dimensional space\cite{busch2023multi,meer2020movie,perl2023low}. Manifold learning is validated to effectively express brain dynamics into a low-dimensional manifold space\cite{gao2015simplicity,noman2024graph,busch2023multi,casanova2021embedding}.

We use UMAP\cite{mcinnes2018umap} to spatially compress the regional time series data, including AWATS and ATS, into a two-dimensional space. In this way, each TR can be represented as a single point in the plane. This approach allows for an intuitive examination of the differences in the patterns of the regional time series data and their separability based on the labels. Moreover, analyzing the visualization results enables us to explore the latent structure of brain dynamics in the low-dimensional space, which is highly significant for a deeper understanding of brain function and mechanisms.

For AWATS, a feature extractor is necessary to extract regional time series from the spatial representation with high dimensionality for each ROI. We adopt the neural network-based feature extractors which are pre-trained in cognitive state decoding. Furthermore, to underscore the importance of the downstream task, we additionally replace the NN with the Principal Component Analysis (PCA) as the feature extractor. Particularly, for each ROI and for each TR, the three representation vectors are concatenated and then decomposed using the PCA, and only the first principal component is retained.

\subsection{Interpretability analysis using Shapley value method}
We perform an interpretability analysis using the Shapley value method to examine the contributions of each ROI in decoding the cognitive states. Through this analysis, we aim to gain insight into how the STDfMRI model utilizes AWATS and ATS of different ROIs. Additionally, this analysis may provide valuable insight into the cognitive processes of the human brain during cognitive tasks.

The Shapley value method is commonly used to estimate the importance of features, particularly in black-box models\cite{sundararajan2020many}. In practical applications, the Shapley value can be approximated using Monte Carlo methods. Detailed information on the accurate derivation and calculation methods can be found in the referenced literature \cite{fan2023model}.

In this paper, we calculate the Shapley values for both AWATS and ATS for further analysis. We carefully examine the influence of the number of simulations. Ultimately, we set the number of simulations as 64 to achieve a balance between computational efficiency and result stability. 

After the calculation of the Shapley values, we obtained the contribution of each ROI in cognitive state decoding. We then perform a visualization for the ROIs’ contribution. For each cognitive state, we obtain contributions of each ROI that aid in identifying the specific cognitive state. To present the relative contribution of each ROI, we normalized the contributions by dividing them by the maximum positive contribution. Subsequently, we generate a brain map using the BrainNet\cite{xia2013brainnet} software, where the surface is color-coded to reflect these normalized contributions.

\section{Results}

\subsection{Cognitive state decoding performance}
We perform cognitive state decoding using both AWATS and ATS to compare their ability to reflect the brain states. We use four metrics to evaluate cognitive state decoding performance including accuracy, precision, recall, and F1 score. We also perform validation in different experiment settings including the variation in resample size, time window size, and training dataset size.

\subsubsection{Comparison between AWATS and ATS at various atlas scales}
Table \ref{tab:atlas_scales} presents a comparison of the accuracy, precision, recall, and F1 score for AWATS and ATS at different atlas sacles. In the comparison experiment, the time window size is set to 15 TRs and the resampling size is set to 10. The results show that AWATS significantly outperforms ATS across different atlas scales (p\textless0.001). This indicates the advantage of AWATS in cognitive state decoding.

\begin{table*}[ht]
  \centering
  \caption{Comparison of the four metrics including accuracy, precision, recall, and F1 score between AWATS and ATS using Schaefer atlases with different number of ROIs. The time window is 15 TRs. T-test is performed to evaluate the significance, which demonstrates that the performance of AWATS significantly better than ATS across different atlas scales (p<0.001) .}
  \begin{tabular}{ccccccccc}
    \toprule
    Methods & \multicolumn{4}{c}{AWATS} & \multicolumn{4}{c}{ATS} \\
    \cmidrule(lr){1-1} \cmidrule(lr){2-5} \cmidrule(lr){6-9}
    Number of ROIs &100&200&300&400&100&200&300&400 \\
    \midrule

    Accuracy & 92.71$\pm$0.3 & 92.94$\pm$0.2 & \textbf{93.51$\pm$0.4} & 92.67$\pm$0.6 & 85.03$\pm$1.2 & 87.22$\pm$0.8 & 88.32$\pm$0.8 & 88.41$\pm$0.7 \\
    Precision & 91.22$\pm$0.8 & 92.37$\pm$0.2 & \textbf{92.46$\pm$1.1} & 91.59$\pm$0.9 & 82.79$\pm$1.4 & 84.87$\pm$0.9 & 86.28$\pm$1.2 & 87.28$\pm$0.7 \\
    Recall & 91.27$\pm$0.6 & 92.36$\pm$0.3 & \textbf{92.79$\pm$1.0} & 91.09$\pm$1.1 & 82.21$\pm$1.2 & 85.19$\pm$0.9 & 86.41$\pm$0.6 & 86.30$\pm$1.0 \\
    F1 score & 91.15$\pm$0.6 & 92.31$\pm$0.2 & \textbf{92.60$\pm$1.0} & 91.26$\pm$1.0 & 82.39$\pm$1.3 & 84.93$\pm$1.0 & 86.23$\pm$0.8 & 86.73$\pm$0.9 \\

    \bottomrule
  \end{tabular}
  \label{tab:atlas_scales}
\end{table*}

\subsubsection{Comparison with voxel-wise methods}
To illustrate the effectiveness of spatial averaging, we compare our AWATS method with two voxel-wise brain decoding approaches: the searchlight and the 3D-CNN. Our implementation of the searchlight method follows\cite{cheung2023decoding} and the 3D-CNN is based on\cite{vu2020fmri}. The searchlight method employs a small spherical or cubical mask that scans across the brain volume, measuring the prediction performance of the corresponding voxels. Subsequently, voxels demonstrating high performance are selected for the final brain decoding test. The 3D-CNN method utilizes 3D convolutional layers to extract features from the entire brain volume and fully connected layers to map these features to predictions.

Table \ref{tab:voxel_wise_methods} presents the comparison results between AWATS, searchlight, and 3D-CNN. For the AWATS method, the Schaefer 300 atlas is used. The time window size is set to 15 TRs and the resampling size is set to 10.The results show that the performance of AWATS is significantly better than the searchlight (p\textless0.001) and the 3D-CNN (p\textless0.001) methods.

\begin{table}[ht]
  \centering
  \caption{Comparison of the four metrics including accuracy, precision, recall, and F1 score between AWATS and voxel-wise methods. Schaefer atlas with 100 ROIs is used and the length of the time window is 15 TRs. T-test is performed to evaluate the significance, which demonstrates that the performance of AWATS significantly better than Searchlight (p<0.001) and 3D-CNN (p<0.001).}
  \begin{tabular}{cccc}
    \toprule
    Methods & AWATS & Searchlight & 3D-CNN \\
    \midrule
    Accuracy & \textbf{93.51$\pm$0.4} & 85.58$\pm$0.6 & 91.54$\pm$0.3 \\
    Precision & \textbf{92.46$\pm$1.1} & 83.40$\pm$0.8 & 91.44$\pm$0.3 \\
    Recall & \textbf{92.79$\pm$1.0} & 83.76$\pm$0.4 & 91.85$\pm$0.3 \\
    F1 score & \textbf{92.60$\pm$1.0} & 83.70$\pm$0.4 & 91.27$\pm$0.2 \\
    \bottomrule
  \end{tabular}
  \label{tab:voxel_wise_methods}
\end{table}

\subsubsection{Robustness of cognitive state decoding}
In order to investigate the robustness of cognitive state decoding using AWATS, we conduct experiments to examine the impact of different time window sizes, resampling sizes, and training dataset sizes on the performance of the model. First, we evaluate the performance of the time window sizes ranging from 3 TRs to 15 TRs, with a step of 2 TRs. Cognitive state decoding accuracies of AWATS and ATS for each window size are presented in Table \ref{tab:window_size}. Both AWATS and ATS exhibit the highest performance when the window size is set to 15 TRs. Notably, AWATS significantly outperforms ATS across different time window sizes and atlas scales in terms of classification accuracy (p\textless0.001). The results demonstrate that the STDfMRI model, utilizing AWATS as input, consistently achieves stable performance when the length of the sliding window is greater than 7 TRs.

Then, we reduce the resampling size when extracting AWATS, a smaller sample size means the feature extractor learns to extract useful information from a coarser spatial representation of voxels within the brain region. We evaluate the performance of different resampling sizes ranging from 2 to 10, with a step of 2. The results for different resample sizes are presented in Table \ref{tab:resample_size}. As the resample size is decreased, the accuracy achieved by the STDfMRI model utilizing AWATS exhibits a slight decline. With a sample size set to 2, AWATS achieves an accuracy of 90.39\% for Schaefer-100 atlas which is higher than 85.03\% achieved by ATS. This indicates the model’s performance remains better than ATS, even when the resample size is set as low as 2.

Finally, we perform an experiment that gradually reduces the size of the training dataset. The impact of the training dataset sizes is evaluated by setting the proportion of the training dataset ranging from 0.3 to 0.8, with a step of 0.1. The accuracies achieved by the model utilizing AWATS and ATS under different training dataset sizes are present in Table \ref{tab:training_dataset_size}. The results show that the model’s performance tends to decrease when a smaller training dataset is used for both AWATS and ATS. Despite this decrease in performance due to the limited amount of training data, AWATS significantly outperforms ATS across different training dataset sizes (p\textless0.001).

\begin{table*}[ht]
  \centering
  \caption{Cognitive state decoding accuracy for different time window sizes. T-test is performed to evaluate the significance, which demonstrates that the performance of AWATS significantly better than ATS across different time window sizes (p<0.001) .}
  \begin{tabular}{ccccccccc}
    \toprule
    \multirow{2}*{Method} & \multirow{2}*{Number of ROIs} &  \multicolumn{7}{c}{Time window size} \\ 
    \cmidrule{3-9}
    & &3&5&7&9&11&13&15\\
    \midrule
    \multirow{4}*{AWATS} &
    100 & 87.74$\pm$1.0 & 88.41$\pm$0.2 & 90.46$\pm$0.8 & 90.20$\pm$1.1 & 90.73$\pm$0.4 & 91.34$\pm$0.6 & 92.71$\pm$0.3 \\
     & 200 & 88.74$\pm$0.4 & 89.76$\pm$1.2 & 91.01$\pm$0.7 & 90.75$\pm$1.0 & 92.12$\pm$0.9 & 91.98$\pm$0.4 & 92.94$\pm$0.2 \\
     & 300 &89.25$\pm$0.9 & 90.37$\pm$0.7 & 91.39$\pm$0.6 & 91.56$\pm$1.1 & 92.62$\pm$0.4 & 92.43$\pm$0.8 & 93.51$\pm$0.4 \\
     & 400 & 88.65$\pm$0.8 & 90.48$\pm$0.9 & 90.82$\pm$1.1 & 92.18$\pm$0.6 & 92.12$\pm$0.9 & 92.91$\pm$0.6 & 93.37$\pm$0.2 \\

    \midrule
    \multirow{4}*{ATS} &    
    100 & 71.54$\pm$1.0 & 76.70$\pm$1.2 & 79.22$\pm$1.3 & 81.56$\pm$1.7 & 82.60$\pm$1.7 & 83.35$\pm$1.0 & 85.03$\pm$1.2 \\
    & 200 & 73.42$\pm$0.9 & 79.20$\pm$0.5 & 81.06$\pm$1.3 & 82.80$\pm$1.6 & 84.26$\pm$1.4 & 85.14$\pm$0.9 & 87.22$\pm$0.8 \\
    & 300 & 76.18$\pm$1.2 & 80.04$\pm$1.2 & 83.31$\pm$1.3 & 84.32$\pm$0.7 & 85.69$\pm$0.7 & 86.82$\pm$0.9 & 88.32$\pm$0.8 \\
   & 400 & 77.39$\pm$1.8 & 80.73$\pm$1.3 & 83.42$\pm$1.4 & 84.90$\pm$0.6 & 87.02$\pm$0.3 & 87.85$\pm$0.4 & 88.41$\pm$0.7 \\

    \bottomrule
  \end{tabular}
  \label{tab:window_size}
\end{table*}

\begin{table*}[ht]
  \centering
  \caption{Cognitive state decoding accuracy of AWATS for different resample sizes.}
  \begin{tabular}{cccccc}
    \toprule
    \multirow{2}*{Number of ROIs} & 
    \multicolumn{5}{c}{Resample size} \\ \cmidrule(lr){2-6}
    & 2 & 4 & 6 & 8 & 10 \\
    \midrule
    100 & 90.39$\pm$0.3 & 91.28$\pm$0.2 & 92.42$\pm$0.4 & 92.65$\pm$0.5 & 92.71$\pm$0.3 \\
    200 & 90.66$\pm$0.5 & 91.52$\pm$0.2 & 92.61$\pm$0.5 & 92.81$\pm$0.2 & 92.94$\pm$0.2 \\
    300 & 91.96$\pm$0.3 & 92.86$\pm$0.3 & 92.98$\pm$0.3 & 93.19$\pm$0.2 & 93.51$\pm$0.4 \\
    400 & 91.81$\pm$0.4 & 92.74$\pm$0.4 & 92.83$\pm$0.5 & 93.11$\pm$0.4 & 93.37$\pm$0.2 \\

    \bottomrule
  \end{tabular}
  \label{tab:resample_size}
\end{table*}

\begin{table*}[ht]
  \centering
  \caption{Cognitive state decoding accuracy for different training dataset sizes.T-test is performed to evaluate the significance, which demonstrates that the performance of AWATS significantly better than ATS across different training dataset sizes (p<0.001) .} 
  \begin{tabular}{cccccccc}
    \toprule
    \multirow{2}*{Method} & \multirow{2}*{Number of ROIs} &  \multicolumn{5}{c}{Training dataset ratio} \\ 
    \cmidrule(lr){3-8}
    & & 30\% & 40\% & 50\% & 60\% & 70\% & 80\% \\
    \midrule
    \multirow{4}*{AWATS} &
    100 & 88.38$\pm$1.0 & 90.11$\pm$0.7 & 90.73$\pm$0.9 & 92.14$\pm$0.8 & 91.87$\pm$0.7 & 92.71$\pm$0.3 \\
     & 200 & 89.64$\pm$0.6 & 90.97$\pm$0.3 & 91.26$\pm$0.7 & 92.56$\pm$0.8 & 92.87$\pm$0.7 & 92.94$\pm$0.2 \\
     & 300 & 89.22$\pm$0.9 & 91.46$\pm$0.8 & 92.62$\pm$0.5 & 93.02$\pm$0.4 & 93.31$\pm$0.5 & 93.51$\pm$0.4 \\
     & 400 & 89.44$\pm$0.8 & 91.39$\pm$0.5 & 92.05$\pm$0.9 & 93.00$\pm$0.3 & 92.67$\pm$0.6 & 93.37$\pm$0.2 \\

    \midrule
    \multirow{4}*{ATS} &    
    100 & 78.94$\pm$0.9 & 80.39$\pm$0.9 & 82.75$\pm$0.9 & 82.98$\pm$1.2 & 84.55$\pm$0.9 & 85.03$\pm$1.2 \\
    & 200 & 81.59$\pm$1.3 & 83.42$\pm$1.2 & 85.41$\pm$0.8 & 84.66$\pm$1.4 & 85.74$\pm$0.8 & 87.22$\pm$0.8 \\
    & 300 & 84.06$\pm$1.1 & 85.40$\pm$1.0 & 86.18$\pm$1.1 & 86.69$\pm$0.8 & 87.31$\pm$0.7 & 88.32$\pm$0.8 \\
   & 400 & 83.31$\pm$0.6 & 84.52$\pm$1.2 & 87.00$\pm$1.1 & 87.10$\pm$1.0 & 87.68$\pm$0.8 & 88.41$\pm$0.7 \\

    \bottomrule
  \end{tabular}
  \label{tab:training_dataset_size}
\end{table*}

\subsection{Visualization analysis based on manifold learning}
We employ the UMAP technique to project AWATS and ATS extracted from the fMRI data into a two-dimensional space. Specifically, for AWATS, we utilize the neural network-based feature extractor that was jointly trained with the cognitive state decoding (AWATS-NN). To provide a comparison, we also use PCA as the feature extraction method (AWATS-PCA). In this section, we present the visualization result of the WM task due to its maximum number of time points. We randomly select 15 subjects for the visualization analysis to ensure the clarity of the figure illustration. The embeddings are normalized to have a mean of 0 and a standard deviation of 1 for an appropriate presentation.

\begin{figure*}[ht]
\centering
\subfloat[]{\includegraphics[scale=0.35]{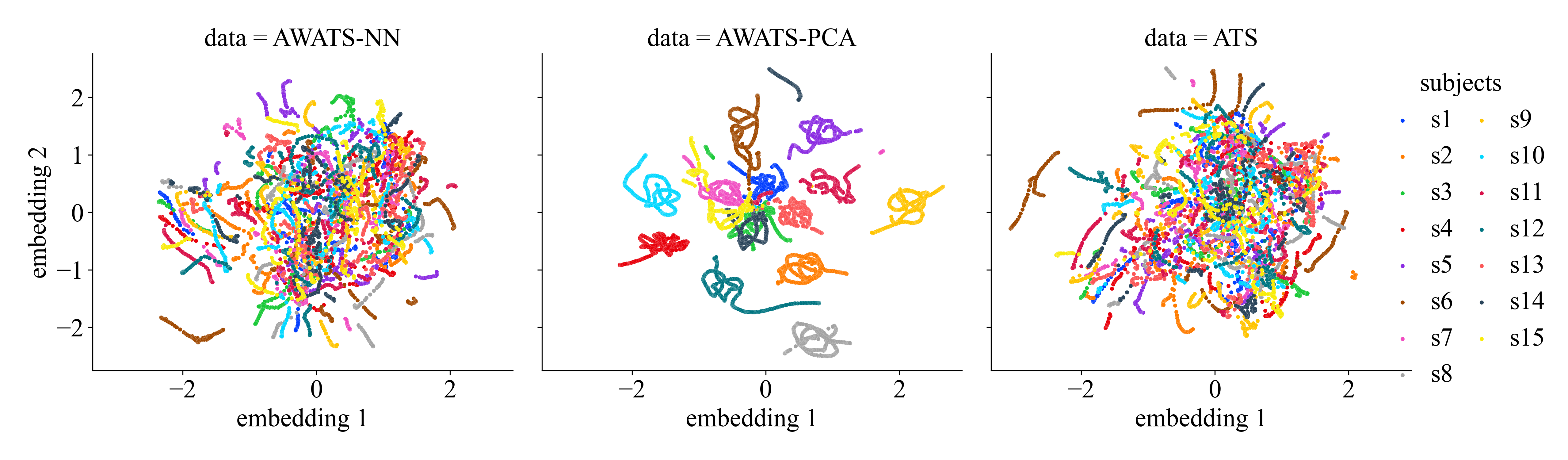}}
\hfill
\subfloat[]{\includegraphics[scale=0.35]{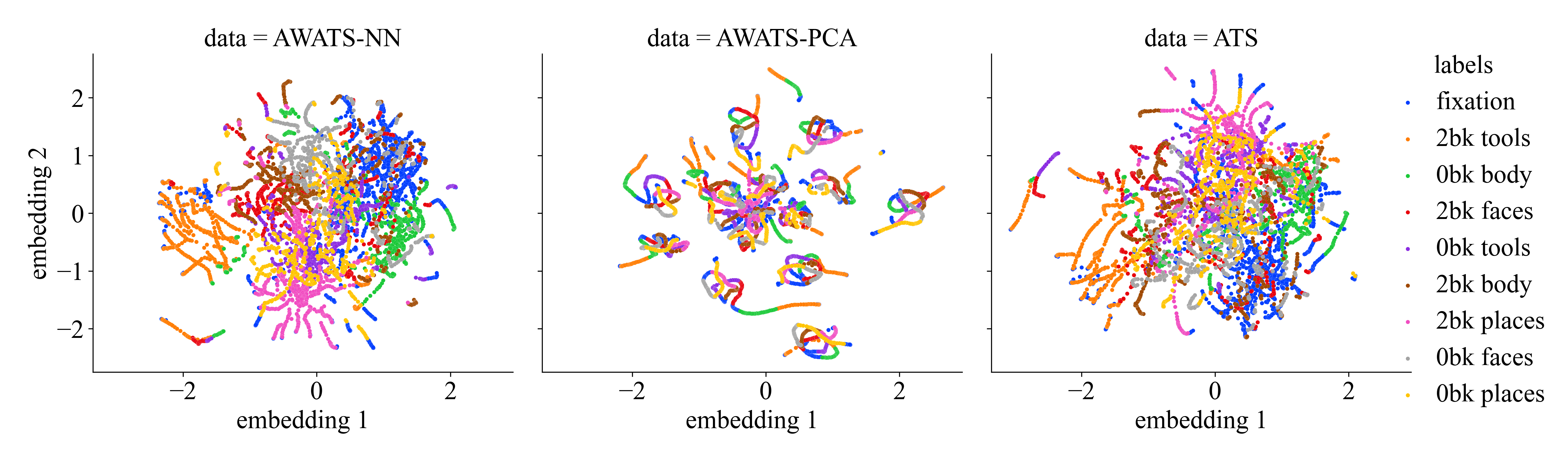}}
\caption{Visualization of the regional time series embedding obtained by UMAP for the WM task. (a) The color of the scatters represents different subjects. (b) The color of the scatters represents different task conditions.}
\label{fig:embedding}
\end{figure*}

The visualization results are shown in Figure \ref{fig:embedding}. We employed two coloring schemes for the embedding visualization. The first (Figure \ref{fig:embedding} (a)) involves assigning the same color to the data from the same subjects, while the second (Figure \ref{fig:embedding} (b)) involves assigning the same color to the data corresponding to the same cognitive state. Figure \ref{fig:embedding} (a) presents how the data from different subjects is spread out in the embedding space. AWATS-PCA stands out for having the clearest separation, while AWATS-NN shows a lot of mixing among different subjects. Specifically, AWATS-PCA tends to generate discrete temporal trajectories for individual subjects, resulting in negligible overlap among different subjects. This indicates that AWATS-PCA contains the subject-specific information. On the other hand, AWATS-NN and ATS have mixed data from different subjects, but AWATS-NN has a wider range for each subject compared to ATS. This indicates that AWATS-NN retains a considerably low subject-specific information, which is irrelevant to the task of decoding cognitive states.

In the context of examining the distribution of data corresponding to different task conditions within the embedding space (as shown in Figure \ref{fig:embedding} (b)), AWATS-NN shows the best separability, while AWATS-PCA struggles to distinguish between different task conditions. For AWATS-PCA, each subject has an independent temporal trajectory and there is no clear connection among the data from the same task condition but different subjects. In contrast, AWATS-NN reveals distinctly defined boundaries among the data points associated with different task conditions, aligning with its better performance in cognitive state decoding. As for ATS, data corresponding to the same task condition occupies a relatively confined range, but there is still a lot of overlap and mixing.

In order to generalize the results of the visualization analysis (performed on 15 subjects) to all subjects, we additionally performed a quantitative analysis. We employed the ratio of mean inter-class to intra-class distance as a metric for feature separability, where a higher value indicates better separability. The separability of feature embeddings across both task conditions and subjects was assessed. The results are presented in Table \ref{tab:quant_sepra}, demonstrating that AWATS-NN exhibits the highest separability among task conditions, but the lowest separability across subjects. This ensures the generalizability of our visualization-based conclusions to all subjects .

\begin{table}[ht]
  \centering
  \caption{The ratio of mean intra-class to inter-class distance for feature embeddings.}
  \begin{tabular}{cccc}
    \toprule
    \multirow{2}*{Label} &  \multicolumn{3}{c}{Method} \\ 
    \cmidrule(lr){2-4}
    & AWATS-NN & AWATS-PCA & ATS \\
    \midrule
    Task condition &1.26&1.02&1.14 \\
    Subject & 1.03&2.43&1.04 \\
    \bottomrule
  \end{tabular}
  \label{tab:quant_sepra}
\end{table}

\subsection{Interpretability analysis based on Shapley value}
\begin{figure*}[ht]
\centering
\subfloat[]{\includegraphics[scale=1.7]{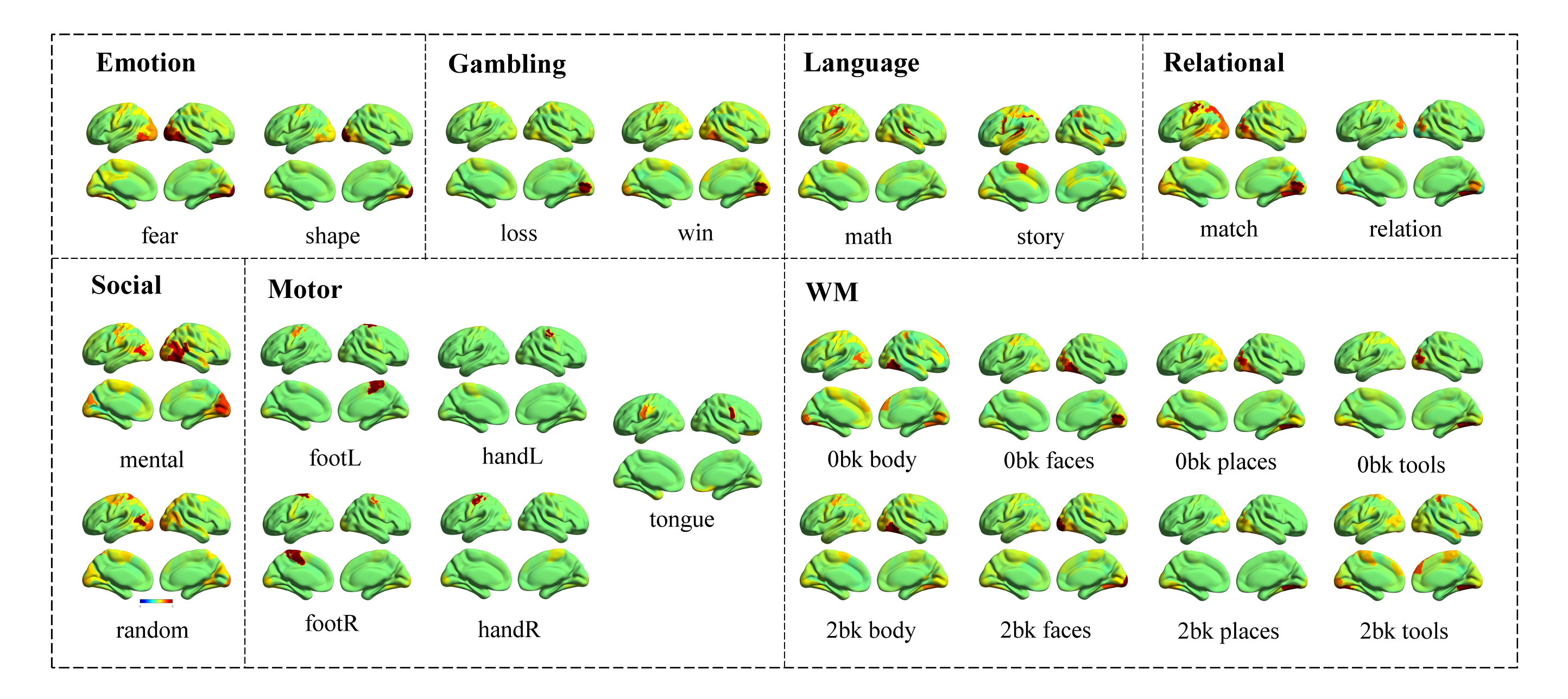}}
\hfill
\subfloat[]{\includegraphics[scale=1.7]{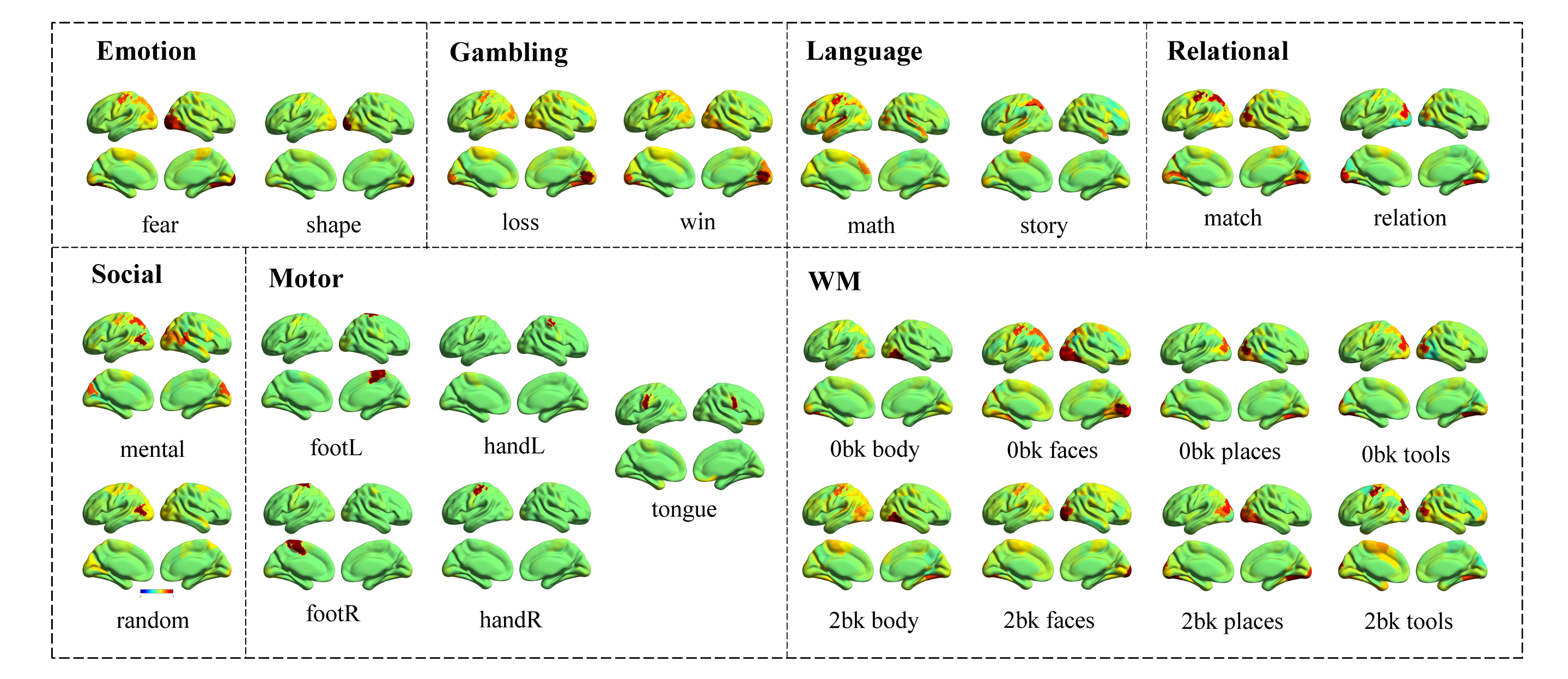}}
\caption{Contribution maps calculated using the Shapley value method. (a) Contribution maps for AWATS. (b) Contribution maps for ATS. The contributions have been normalized by dividing them by the maximum positive contribution to depict the relative importance of different brain areas in identifying specific task conditions.}
\label{fig:contribution_map}
\end{figure*}

We calculate the Shapley values to access the contribution of each ROI for both AWATS and ATS using the Schaefer-100 atlas. Subsequently, we visualized the relative contributions of different ROIs to each cognitive state, as shown in Figure \ref{fig:contribution_map}. It’s worthing that the contribution maps of AWATS and ATS exhibit similar pattern. However, on a broader scale, when using AWATS, the model tends to concentrate more on specific areas and exhibits a smaller number of high-contribution areas on the contribution map.

In the language task, high contributions are identified in the superior temporal gyrus (STG) and the inferior parietal lobule, which are known to correlate with language processing\cite{acheson2013stimulating,binkofski2016neuroanatomy,yi2019encoding}. Besides, high contributions are identified in the sensorimotor cortex including the precentral gyrus and the supplementary motor area (SMA). For ATS, other high contribution areas also recognized including the middle frontal gyrus (MFG) and the superior frontal gyrus (SFG).

In the relational task, high-contributions are found in the inferior parietal lobule, the sensorimotor cortex, and the visual cortex. Compared to ATS, AWATS highlights the MTG, which is associated with accessing word meanings while reading\cite{acheson2013stimulating} and may contribute to understanding the relations between words.

In the motor task, the high-contribution areas identified include precentral gyrus (corresponding to the primary motor cortex\cite{roux2020functional}), postcentral gyrus (corresponding to the primary somatosensory cortex\cite{zhou2020diffusion}), paracentral lobule (corresponding to motor and sensory functions of the lower extremity\cite{patra2021morphology}), and supplementary motor area (SMA), contributing to movement control. 

In the WM task, the contribution maps of both AWATS and ATS highlight the combined effect of prefrontal lobule, temporal lobule, and occipital lobule. Common high-contribution areas for both AWATS and ATS are the inferior occipital gyrus, fusiform gyrus, and medial frontal gyrus. However, distinctions are observed. Compared to AWATS, ATS places a stronger emphasis on the sensorimotor cortex, including the precentral gyrus, postcentral gyrus, and SMA.

To validate the results of the Shapley value method, we also conduct interpretability analysis using logistic regression for both AWATS and ATS. As a form of linear classification model, logistic regression offers better interpretability compared to deep neural networks. Specifically, we fit logistic regression models for AWATS and ATS to predict cognitive states, considering the coefficient for each feature as its contribution. The contribution maps based on logistic regression are presented in Figure S1. Although contribution maps based on logistic regression and Shapley value exhibit discrepancies due to the different principles of interpretation, the results from both methods are generally consistent, especially for the emotion, social, and motor tasks.

To investigate the contributions of the brain network activities in identifying cognitive state, we computed the averaged contributions for the Yeo's 7 brain networks\cite{yeo2011organization}. For AWATS and ATS, statistical analysis  The contributions of these 7 brain networks to the 23 task conditions are presented in Figure S2. Statistical analysis demonstrates that these two methods are not significantly different (p$>$0.05) in network contribution. The network contributions appear similar for both methods. It is shown that SMN plays a crucial role in identifying the motor task conditions, whereas VIS holds more significance for conditions related to other tasks.

\section{Discusstion}
In this study, we propose the AWATS extraction method that, to the best of our knowledge, is the first attempt to incorporate the spatial information within each brain ROI when extracting regional time series from fMRI data. We conduct extensive experiments to validate the advantage of AWATS, over the traditional ATS, covering on cognitive state decoding and manifold learning. First, we propose to use neural networks that are jointly trained with cognitive state decoding to adaptively weight the voxels within each ROI, preserving the spatial information. Then, we designed the STDfMRI model for cognitive state decoding, aiming to map the regional time series of each ROI with different cognitive states. Next, we visualize AWATS and ATS by spatially embedding them into a two-dimensional space and analyze their distribution and separability. Lastly, we perform the interpretability analysis on AWATS and ATS to identify the most informative ROI. Through these analyses, we demonstrate that AWATS retains more task-related information within specific brain regions. This highlights the potential of our proposed method in capturing meaningful patterns in fMRI data and its implications for cognitive neuroscience research.

\subsection{AWATS outperforms ATS in cognitive state decoding}
Identifying cognitive states from task-related fMRI has been the subject of extensive research. Prior studies\cite{vu2020fmri,ye2023explainable} have suggested that the performance of cognitive state decoding is influenced by both the model design and the input features. In this paper, we perform cognitive state decoding using both AWATS and ATS. We propose the STDfMRI model to predict the cognitive states using the regional time series data, which models the brain dynamic process both from the spatial and temporal aspects.

The results reveal that AWATS outperforms ATS in cognitive state decoding regardless of window sizes, resampling sizes, and training dataset sizes, indicating that the task-specific and spatial information captured within each ROI using the AWATS extraction method may be helpful in cognitive state decoding. Additionally, we compare our method with voxel-level methods including the searchlight and  the 3D-CNN methods. The results show that our method achieve better performance in brain decoding, which demonstrates the effectiveness of adaptively averaging.

We compare the performance of cognitive state decoding for AWATS and ATS using the Schaefer atlases with different granularity in brain region parcellation. Consistently, AWATS outperforms ATS for all of the atlases. Interestingly, ATS using atlas with more number of ROIs achieves better performance, which aligns with the prior findings\cite{ye2023explainable}. However, AWATS achieves consistent performance regardless of atlas' granularity. This further underscores our method’s efficiency in cognitive state decoding. This may indicate that our method effectively captures relevant information within each ROI, mitigating information loss in simple voxel averaging. Consequently, a more detailed atlas may not significantly improve the cognitive state decoding performance.

We perform experiments to investigate the robustness of our method across variations in time window sizes, resampling sizes, and training dataset sizes. The results consistently demonstrate our method's advantage over ATS, presenting its robust performance across diverse experimental settings. The results of different time window sizes reflect the extent of temporal information provided to the model of a single task condition, and the results of different training dataset sizes signify the ability of the algorithm that learns the data distribution of the entire dataset from the training dataset. Decreasing the window size or the training dataset size diminishes the performance of both AWATS and ATS. And AWATS consistently outperforms ATS under these conditions. Resampling sizes determine to what extent the details in the spatial representations for each ROI can be retained. Interestingly, reducing the resample size from 10 to 2 incurs a smaller decrease in cognitive state decoding performance compared to reducing the time window size and training dataset ratio. This suggests the effectiveness of extracting the spatial information of our method, even when describing spatial distribution within the ROI using the representation vectors with a smaller size.

\subsection{AWATS achieves better separability of task conditions in embedding space}
There is a prevailing belief that intricate brain dynamics may manifest in a conceivable low-dimensional space. Several studies\cite{bahrami2019using,casanova2021embedding,gao2020non} have substantiated the feasibility of leveraging non-linear manifold learning techniques to unveil the trajectory of brain states in latent space. In this paper, we utilize the manifold learning method UMAP to embed fMRI regional time series data into a two-dimensional space. In the embedding space, we carefully compare the distributions and separabilities of the embedded AWATS and ATS.

In this study, two kinds of feature extractors are employed to extract AWATS. First, we adopt the pre-trained feature extractor from cognitive state decoding (AWATS-NN). Meanwhile, we also intend to investigate the effect of the information of downstream tasks. Thus, we introduce a PCA-based feature extractor (AWATS-PCA) that retains the first principal component for each ROI.

Upon examination across various subjects, AWATS-PCA exhibits significantly better separability among different subjects than AWATS-NN and ATS, suggesting that PCA effectively retains the task-irrelevant individual information.  In fact, AWATS-PCA reflects the main pattern of the spatial representations generated using our method, while AWATS-NN contains information adaptively extracted from spatial representations by jointly training with the cognitive state decoder, i.e. the STDfMRI model. The shift in distribution and separability in the embedding space from AWATS-PCA to AWATS-NN indicates that the feature jointly trained in cognitive state decoding effectively neglects the subject-specific information and extracts relevant information for the downstream task.

\subsection{AWATS identified reasonable high-contribution brain areas}
Interpretability analysis is instrumental in understanding why our method achieves better performance in cognitive state decoding and in identifying the brain regions that are specially activated when performing one task compared to another task or the resting state. In this study, we perform a Shapley value-based interpretability analysis. The interpretability analysis underscores the potential of the AWATS in providing valuable insights into brain functions and cognitive processes.

The contribution maps for AWATS and ATS exhibit similar patterns, and the finding is physiologically reasonable. The similarity of the contribution maps between AWATS and ATS suggests that adaptive weighting using neural networks in regional time series extraction does not compromise interpretability. Common high-contribution brain areas are identified, including the sensorimotor cortex for the motor task and the visual cortex for other cognitive tasks. The sensorimotor cortex, encompassing the primary motor cortex, somatosensory cortex, and supplementary motor cortex, is thought to be the core region for human motions\cite{borich2015understanding,penfield1937somatic}. The visual cortex is consistently identified as playing a key role in contributing to cognitive tasks, excluding motor task. This observation may stem from the varied visual stimuli presented in the task conditions. Both the sensorimotor cortex and the visual cortex predominantly process perceptual information with a single modality\cite{baum2020development}. This characteristic may render information from these regions easily understandable by the STDfMRI model.

In this paper, we perform cognitive state decoding using deep neural networks, which present challenges in terms of interpretability. To address this, we conduct an interpretability analysis based on Shapley values, treating the models as black boxes and inferring the contribution of ROI based on prediction performance. However, the interpretability of deep neural networks in cognitive state decoding remains a topic of controversy\cite{thomas2023benchmarking}. Therefore, we also performed an interpretability analysis using logistic regression, where the fitted coefficients are considered as feature importance indicators. The results show that the interpretability analysis of the Shapley value method and logistic regression identified contribution maps in a similar pattern. This finding validates the effectiveness of the Shapley value method for the STDfMRI model. Additionally, differences between the results of these two methods may be attributed to their distinct principles of explainability and the limitations of the decoding ability of logistic regression.

\subsection{Limitations and future work}
In this paper, we have employed cognitive state decoding as the task to validate the ability of our method to provide valuable spatial information. However, there are numerous other tasks that can be utilized for validating our method, such as disease diagnosis and human phenotype prediction. These tasks can also be used to assess the effectiveness of our approach. In future work, we plan to conduct experiments on various tasks to further validate the general applicability of our method. Moreover, the generalization ability of the deep learning-based method should be further validated on different datasets, although the network should be trained from scratch on each dataset.

\section{Conclusion}
In this study, we consider the basic question during fMRI-based brain region identification for cognitive states, i.e. how to extract 1D regional time series data that can represent the functional fluctuations of each ROI. We propose to use neural networks that are jointly trained with the cognitive state decoder to adaptively retain the spatial information among voxels within a ROI. We have performed extensive experiments to evaluate the benefits of our method over the traditional averaging method. In cognitive state decoding, AWATS consistently outperforms ATS across various experiment settings, i.e. different window sizes, resampling sizes, and training dataset sizes. The manifold learning-based visualization analysis demonstrates that the proposed adaptively weighted spatial averaging method effectively neglects the subject-specific information and achieves enhanced separability of task conditions compared to ATS. Moreover, the results of interpretability analysis results indicate that AWATS identifies reasonable high-contribution areas, suggesting it captures meaningful information. These findings demonstrate the effectiveness of our method in preserving more informative details of the functional fluctuations of brain regions and its potentially benefit in identifying high-contribution brain regions in various cognitive states. Our work regional time series may aid the improvement of the basic pipeline of fMRI processing.

\section*{Acknowledgements}
This work is founded by National Natural Science Foundation of China (No.62076080, 62306083), Natural Science Foundation of ChongQing CSTB2022NSCQ-MSX0922 and the Postdoctoral Science Foundation of Heilongjiang Province of China (LBH-Z22175).

\bibliographystyle{unsrt}  
\bibliography{mycite}

\end{document}